\documentclass[traditabstract]{aa}

\usepackage{graphicx}
\usepackage{txfonts}
\usepackage{multirow}
\usepackage{longtable}
\usepackage{color}

\begin{document}

\newcommand{\kms}{km\,s$^{-1}$}
\newcommand{\um}{\mu}
\newcommand{\rstar}{R$_{\star}$}
\newcommand{\msun}{M$_{\odot}$}

\title{On massive dust clumps in the envelope\\ of the red supergiant VY Canis Majoris}

\author{T. Kami\'nski\inst{\thanks{Submillimeter Array Fellow}}
        }

  \institute{\centering Center for Astrophysics \textbar \ Harvard \& Smithsonian,  Smithsonian Astrophysical Observatory, \email{tkaminsk@cfa.harvard.edu}
       }
            
  \date{Received; accepted}

\abstract{The envelope of the red supergiant VY\,CMa has long been considered an extreme example of episodic mass loss that is possibly taking place in other cool and massive evolved stars. Recent (sub-)millimeter observations of the envelope revealed the presence of massive dusty clumps within 800\,mas from the star which reinforce the picture of drastic mass-loss phenomena in VY\,CMa. We present new ALMA observations at an  angular resolution of 0.1\arcsec and at an unprecedented sensitivity that reveal further details about the dusty clumps. We resolve more discrete features and identify a submillimeter counterpart of a more distant Clump\,SW known from visual observations. The brightest clump, named C, is marginally resolved in the observations. Gas seen against the resolved continuum emission of clump\,C produces a molecular spectrum in absorption, in lines of mainly sulfur-bearing species. Except for SW\,Clump, no molecular emission is found to be associated with the dusty clumps and we propose that the dusty structures have an atypically low gas content. We attempt to reproduce the properties of the dusty clumps through three-dimensional radiative-transfer modeling. Although a clump configuration explaining the observations is found, it is not unique. A very high optical depth of all clumps to the stellar radiation make the modeling very challenging and requires unrealistically high dust masses for one of them. It is  suggested that the dusty features have substructures, e.g. porosity, that allows deeper penetration of stellar photons within the clumps than in a homogeneous configuration. A comparison of the estimated clumps ages to variations in the stellar visual flux for over a century suggests that the mechanism responsible for their formation is not uniquely manifested by enhanced or strongly diminished visual light. The study demonstrates that the dusty mass-loss episodes of VY\,CMa are indeed unparalleled among all known non-explosive stars. The origin of these episodes remains an unsolved problem.}
 

\authorrunning{Kami\'{n}ski et al.}
\titlerunning{The dust clumps of VY\,CMa}
\maketitle

\section{Introduction}
Observations of the envelope of the red supergiant VY\,CMa evidence mass loss that have been taking place for over $\gtrsim$1200\,yr and at an average rate of 6$\cdot$10$^{-4}$\,M$_{\sun}$ yr$^{-1}$ \citep{Shenoy2016}. It is estimated that the total envelope mass traced in continuum emission is at least 0.7\,M$_{\sun}$ \citep{Shenoy2016}. Such a high mass-loss rate critically determines the future evolution of this massive star but it is unclear how long the star can sustain this mass loss at its current evolutionary stage. Our understanding of the mass-loss mechanism in red supergiants -- and in VY CMa in particular -- is possibly poorest among all cool evolved stars. Evolutionary modelers often assume crudely calibrated mass-loss formula \citep[e.g][]{deJagger1988} and consequently mass loss is considered a major source of uncertainty in theoretical astrophysics \citep[e.g.][]{GE2015}. VY\,CMa constitutes an unparalleled case of supergiant mass loss in the Galaxy and as such offers a chance to identify physical processes responsible for the most extreme mass loss in non-explosive stars. 

The envelope of VY\,CMa is particularly inhomogeneous. Observed at sufficient angular resolutions, the envelope has revealed a wide diversity of structures, e.g. knots, arcs, clumps, rings,  at scales ranging from mas to arcsec, and in all accessible wavelength regimes, from visual to radio  \citep{Smith2001,H2005, H2007,kamiSurv,DecinNaCl,anita,OGorman,gordon}. Their structure and distibution is very erratic and the envelope does not appear to have any preferred axis of symmetry. Some of the discrete features indicate mass-loss episodes at rates as high as 10$^{-2}$--10$^{-3}$ M$_{\sun}$\,yr$^{-1}$. Their origin is unclear but some studies suggest a link to activity at the stellar surface that involves convection or pulsation or magnetic fields \citep{Smith2001,H2005,H2007,OGorman,Vlemmings}. VY\,CMa may be the best known source to display such episodic mass loss but other Galactic red supergiants have highly structured outflows too \citep[e.g.][]{mauron,noema}. There may be also extragalactic analogs --- clumpy structures in winds of red supergiants that are immediate progenitors of core-collapse supernovae (SNe) give rise to particular observational characteristics of the explosions, including radio and X-ray emission. Spectral energy distributions and time evolution of such  supernovae are strongly influenced by substructures in the progenitors' dusty and H-rich circumstellar envelopes \citep{ChevalierFransson}. Some of such interacting supernova progenitors require mass-loss rates of 10$^{-6}$--10$^{-4}$ M$_{\sun}$\,yr$^{-1}$ \citep{ChevalierFransson} [or even 10$^{-3}$--10$^{-1}$ M$_{\sun}$\,yr$^{-1}$ in case of progenitors of Type IIn SNe \citep{chandra}]. These dusty envelopes could have been ejected in a similar mechanism as that responsible for the mass loss we witness in much greater detail in VY\,CMa. With an initial mass of  $\geq$25\,M$_{\sun}$ \citep{wittkowski}, it is however uncertain whether VY\,CMa itself will go off as a supernova in the red supergiant stage or collapse directly into a black hole or evolve into a hotter star \citep[e.g.][]{Davies} -- with its mass loss being the main source of the uncertainty.  

The advent of the Atacama Large Millimeter/submillimeter Array (ALMA) allowed to resolve the innermost 1\arcsec\ region around VY\,CMa revealing dusty structures on spatial scales 70--700\,AU from the star \citep{anita,OGorman,Vlemmings}. They contain mainly cool (<100\,K) and optically thick dust with unknown contribution of atomic or molecular gas. The brightest structure observed, known as clump C, was estimated to have 1.2$\cdot$10$^{-3}$\,M$_{\sun}$ of dust \citep{Vlemmings} but owing to high optical depth the actual mass may be much higher than this. Clump C and similar structures found in ALMA continuum maps represent relatively recent mass-loss episodes and it is of prime interest to characterize their properties and origin. Here we investigate the structure of the dusty envelope of VY\,CMa based on ALMA observations at sensitivity a few times better than in similar earlier studies \citep{kamiSurv,anita,OGorman}. We critically examine the mass constraints on the dusty medium finding that its mass might have been considerably underestimated in earlier studies, making the VY\,CMa's envelope similar to some progenitors of of supernovae interacting with particularly dense circumstellar environments.

\section{Observations}\label{sec-obs}
We report ALMA observations obtained in Bands 6 and 7 within a project 2013.1.00156.S (PI: T. Kami\'nski). The observations were originally designed to study aluminum- and titanium-bearing species in VY\,CMa \citep{KamiVienna}. The spectral coverage was optimized to cover multiple transitions of these species. Older observations from the Science Verification (SV) phase of ALMA and obtained in Bands 7, 9 \citep{anita,OGorman}, and 5 \citep{Vlemmings} have lower sensitivity and a typically slightly lower angular resolution than our data.

Band\,6 observations took place on 27 Sept. 2015 with 40\,min spent on source. The observations covered the spectral ranges 220.6--224.3, 235.2--237.1, 238.1--240.0\,GHz at a channel binning of about 1.3\,\kms. We used the main array with 34 functioning antennas at projected baselines of 33--2193\,m. The configuration resulted in a beam FWHM of 173$\times$133\,mas and with the major axis at a position angle (PA) of 70\degr. The maximum recoverable angular scale (MAS) of 0\farcs7 is much smaller than the entire field of view of 27\farcs5. 

Band\,7 observations were acquired in two blocks executed on 28 June and 27 September 2015 with 34 and 37 functional antennas (45 different antenna pads). The combined data are equivalent to 71\,min integration on source. The spectral setup covered four ranges: 343.6--345.4, 345.5--347.2, 355.5--357.3, 357.3--359.2\,GHz at a channel width of 1.7\,\kms. Projected baselines of 42--2257\,m combined gave a synthesized beam of 115$\times$98\,mas at a PA of 61\degr\ and a MAS of $\sim$0\farcs9. The field of view at these frequencies is of 18\arcsec.   

All the 2015 observations were initially calibrated with the default CASA pipeline in version 4.3.1. The calibration of the bandpass and complex gains were imperfect. In particular, sub-optimal phase transfer from three different gain calibrators resulted in different source positions in each of the observations, with largest discrepancy of $\sim$56\,mas between Band 6 and 7 data. The offset cannot be explained by the source proper motion which is known with a high accuracy \citep{zhang}. (A similar offset was noticed in the older SV data by \citet{OGorman}.) The offset was determined by measuring positions of brightest continuum features and corrected either in the calibrated visibilities or directly by modifying absolute coordinates in processed images (as indicated in the text below). Both Band 6 and 7 data were also self-calibrated in phase and amplitude. Extra flagging of bad data was introduced in that procedure. The self-calibration greatly improved the signal-to-noise ratios of the data. The final sensitivities of all in-band observations combined were of 27\,$\mu$Jy and 206\,$\mu$Jy at Bands 6 and 7, respectively, as measured at the beam sizes specified above. 

All data were processed in CASA (versions 4.7 to 5.4) and using the {\tt clean} or {\tt tclean} implementations of the CLEAN algorithm. We used the Briggs weighting of visibilities at the robust parameter of 0.5, unless otherwise stated. All velocities here are expressed in the Local Standard of Rest (LSR) frame. Line crowding, especially in Band 7, is severe at our sensitivity levels and it is difficult to find spectral ranges that sample pure continuum. Owing to a complex kinematics and chemical variations of the mm/submm nebula, the windows of pure continuum change very much with position. Standard imaging of continuum was thus not possible at a satisfactory signal-to-noise ratio in this case. Instead, in an iterative procedure, we defined relatively reliable ranges with pure continuum which were then used for a simple linear fit to the visibilities of each spectral window (as defined in ALMA and CASA). These fits were subtracted from the visibilities and used as a representation of the real continuum. In this approach, we took advantage of the continuum flux underlying the numerous spectral lines but underestimate the real noise in the data. 

\begin{figure*}[!ht]
\sidecaption
\includegraphics[angle=0,width=12cm,trim={0 12 0 0},clip]{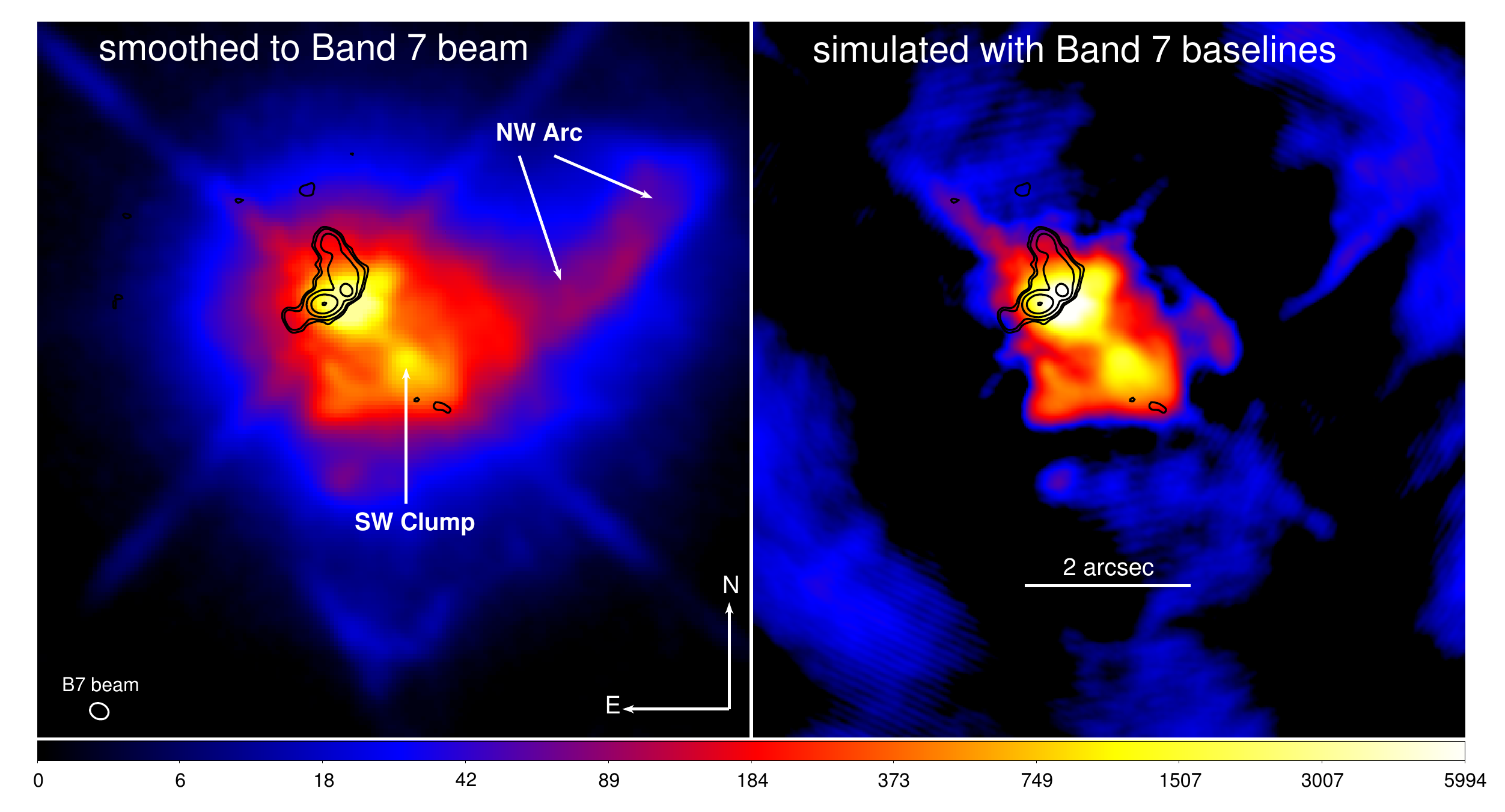}
\caption{Image of the VY\,CMa's nebula from HST/$F1042M$. The image shows mainly scattered light at 1\,$\um$m (and the HST diffraction cross). The left image was smoothed to the Gaussian beam of our Band 7 ALMA observations. The right image was processed with a spatial-frequency filter consistent with the ALMA array configuration used in ALMA observations. Largest features, e.g. NW\,Arc, are filtered out and some other structures are buried in an irregular pattern introduced by a complex interferometer response function. The colorscale of both images is in counts per second. Black contours show continuum emission observed in Band\,7. The contours are drawn from 0.4\% ($\sim$5$\sigma$) to 97\% of the peak emission with a logarithmic step in each contour.}\label{hst}
\end{figure*}

Our interferometric data have a limited response to the spatial distribution of emission in the sky. To visualize how this limitation could affect our analysis, we simulated how the large-scale nebula of VY\,CMa seen in an infrared image would look like filtered by our ALMA baselines. We used the CASA simulator tasks {\it simobserve} and {\it simanalyze} for that purpose. An archival image obtained in 2009 with the HST in the $F1042M$ band ($\sim$1\,$\mu$m) was converted into visibilities using the actual antenna positions and a range of hour angles used in our Band\,7 observations. These visibilities were next imaged with Briggs robust weighting. No extra noise was added to the data. A comparison of the CLEANed image with the HST original image smoothed to the same angular resolution (Fig.\,\ref{hst}) reveals that large-scale diffuse emission and largest structures, such as NW\,Arc, would be filtered out by the ALMA configurations. In particular, should the visual SW\,Clump have a bright and extended submillimeter counterpart, it would not be detected in the ALMA observations imaged at a standard Briggs weighting. 

We attempted a correction of the limited spatial response of the ALMA configurations to structures larger than $\sim$0\farcs9 by combining the calibrated Band\,7 ALMA visibilities with 2010 survey observations of VY\,CMa obtained with the Submillimeter Array (SMA) in the range 279--355\,GHz \citep{kamiSurv}. The SMA observations used an extended configuration of the SMA with projected baselines between 12 and 202\,m which provided a nominal angular resolution of 0\farcs9 and MAS of about 6\arcsec. They were sensitive to larger spatial scales than the ALMA data but had a $\sim$20 times lower continuum sensitivity in the overlapping spectral range. To partially compensate for the low per-band sensitivity, we used the SMA survey data combined over the full frequency range, 279--355\,GHz. We also included data acquired on 23 and 24 February 2017 with the SMA in the extended configuration with projected baselines of 16--189\,m. These more recent observations used the SWARM correlator and covered 314.0--322.3 and 330.2--338.4\,GHz. The observations were obtained and calibrated in a similar fashion as the survey data. All the SMA data from 2010 and 2017 add up to $\sim$34\,h of on-source time resulting in sensitivity that is however still $\sim$9 times below the sensitivity of the ALMA data. The relative visibility weights of the three datasets were recalculated within CASA based on: the actual noise performance in each dataset; the full spectral coverage; time spent on source; number of baselines; number of polarizations; and surface area of the antennas used. These weights were then re-scaled in imaging procedures to enhance different aspects of the combined data. Before the combination, the ALMA Band\,7 visibilities were shifted in position to match continuum peaks in both observations when imaged at the same angular resolution. 

\section{Results}
\subsection{Main continuum emission features}
\begin{figure*}
\includegraphics[angle=0,width=\textwidth,trim={0 0 0 0},clip]{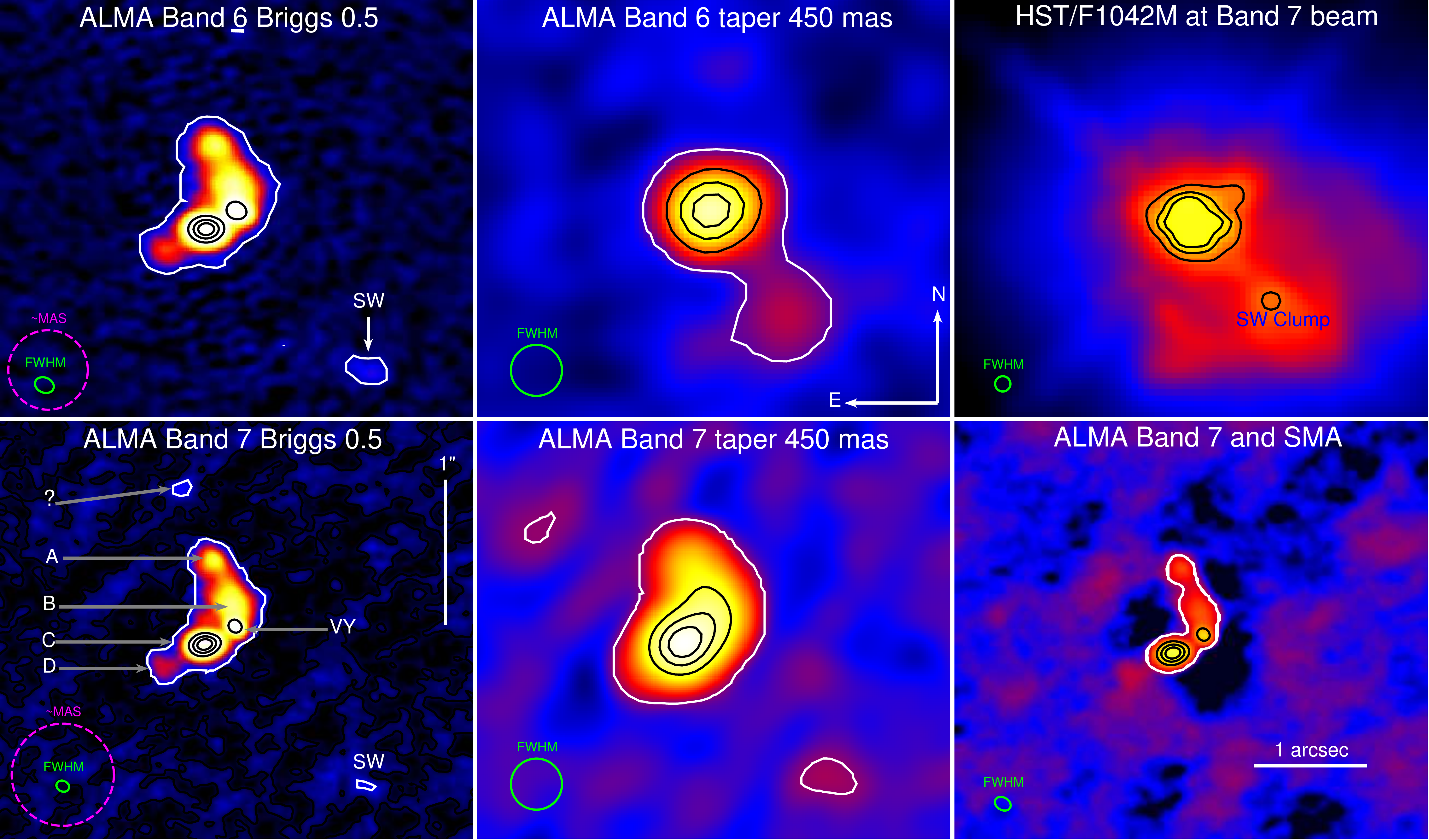}
\caption{Maps of continuum emission of VY\,CMa's dusty envelope. Left panels show maps of emission imaged with ALMA Bands 6 (top) and 7 (bottom) at Briggs weighting (robust 0.5). Middle panels show the corresponding data imaged with a {\it UV} taper which reduced the resolution to a FWHM of 450\,mas. Top right panel shows an HST image at 1\,$\mu$m smoothed to the angular resolution of ALMA Band\,7 observations. Bottom right panel presents a map generated from combined visibilities of ALMA and SMA. The corresponding spatial resolution (FWHM) of each map is shown with green ellipse in lower left corners. In the left panels, we additionally show the size of the largest angular scales. White countours are drawn at the respective 5$\sigma$ levels. Black contours are shown at 3 levels spaced linearly from the peak to zero intensity (excluding zero).}\label{fig-allMaps}
\end{figure*}

Maps of continuum emission observed with ALMA are shown in Fig.\,\ref{fig-allMaps}. Left panels show the images at the standard Briggs weighting and are reminiscent of maps of \citet{OGorman}. At a closer look, however, the new maps reveal more details and new emission components. We divide the source into five main features which appear as individual discrete emission regions, or clumps. Hereafter, we call these features from north to south A, B, VY, C, D, as indicated in Figs.\,\ref{fig-allMaps}--\ref{fig-spindex}. In this convention, we kept the names of features VY and C introduced by \citet{anita}. Feature VY is consistent with the position of the star. Features A, B, VY appeared as a single elongated structure in data at lower resolution but here feature A appears to be a separate clump. B and VY are connected with a bridge of emission at our resolution but their spectral characteristics is different (below) so we distinguish them as separate clumps. Weak feature D was seen in earlier Band 7 data and \emph{appears} as a faint extension of clump C, which is the brightest region in the submm/mm continuum. We characterized these individual structures -- including their position, size, and flux -- by fitting two-dimensional Gaussian functions to the combined Band 6 and 7 data. The fit was performed sequentially in the order A, C, VY, B, and D; after fitting a single Gaussian the model profile was removed before fitting the next feature. Results of these measurements are presented in Table\,\ref{tab-obs}.   

\begin{table*}
\caption{Measured properties of the continuum clumps observed with ALMA.}\label{tab-obs}
\centering
\small
\begin{tabular}{c cc cc cc ccc cc cc}
\hline
Clump& RA     & Dec    & Projected & Kinematic & Major & Minor & Major  & Flux  & Flux  & \multicolumn{2}{c}{Molecular emission}\\
ID   & offset & offset & distance  & age limit & FWHM  & FWHM  & PA     & in B6 & in B7 & range & peak\\
     & (mas)  & (mas)  & (mas)     & (yr)      & (mas) & (mas) & (\degr)& (mJy) & (mJy) & \multicolumn{2}{c}{(\kms)}\\
\hline
VY	 & 0	  & 0	   & 0	     &$\sim$10& 147$\pm$24 &~~67$\pm$32 &~~14$\pm$10	&~~57.38$\pm$0.08 & ~~63.62$\pm$0.63 &--37:75 & 20\\
A	 & 184.8  & 529.5  &~~560.8	 &$\gtrsim$53	 & 218$\pm$18 & 141$\pm$15 &~~41$\pm$08	&~~18.58$\pm$0.12 &  467.18$\pm$0.53 &--14:55 &~~4\\
B	 &~~50.1  & 234.4  &~~239.7	 &$\gtrsim$23	 & 272$\pm$39 & 133$\pm$28 &~~94$\pm$08	&~~27.66$\pm$0.13 &  149.88$\pm$0.48 &--34:84 & 13\\
C	 & 266.4  &--181.9 &~~322.6	 &$\gtrsim$31	 & 181$\pm$15 &~~97$\pm$14 & 111$\pm$06	& 160.39$\pm$0.10 & ~~91.22$\pm$0.69 &--37:62 & 25\\
D	 & 629.8  &--379.2 &~~735.1	 &$\gtrsim$70	 & 177$\pm$66 &~~91$\pm$55 &~~43$\pm$27	&~~~4.20$\pm$0.10 & ~~12.35$\pm$0.51 &--18:57 & 26\\[5pt]
SW\tablefootmark{a}&--613&--909&1096&$\gtrsim$104& 	          & 	       &            & $\gtrsim$3.16\tablefootmark{a}& &--43:60 & 18\\
\hline
\end{tabular}
\tablefoot{
From left to right the columns present: the clump identification label; fitted central RA and Dec position of the clump with respect to the position measured for clump VY; projected angular distance of the clump from VY; lower limit on the kinematical age of the clump; major and minor axis FWHM sizes; the position angle of the major axis of the fitted Gaussian profile; fluxes and their 1$\sigma$ uncertainties in Bands 6 and 7; range of LSR velocities and peak velocity observed in molecular gas toward the same region. The lower limits of the kinematical age were derived using the projected angular distance and an outflow velocity of 60\,\kms.
\tablefoottext{a}{The SW Clump position and flux were measured in Band\,6 {\it UV}-tapered data (see text).}
}\end{table*}

Figure\,\ref{fig-allMaps} reveals two other emission regions. The leftmost panel presenting Band 7 data shows a feature $\sim$1\farcs4 north from VY and labeled as "?". Although its peak emission is above the map 5$\sigma$ noise level, we consider it only tentatively detected as it is not seen in the more sensitive Band 6 data and no counterpart is seen in the visual or infrared. Another feature is seen in both Band 6 and 7 maps $\sim$1\farcs8 south-west from VY. It appears to be close to the visual feature known as SW\,Clump (cf. Fig.\,\ref{fig-allMaps}) but is located at a slightly larger projected distance, almost as it was at the outer edge of the visual manifestation of the clump (Fig.\,\ref{hst}). As discussed above, our ALMA data alone at a standard Briggs weighting (robust 0.5) are not particularly sensitive to features of the size of SW\,Clump and larger. We thus imaged the ALMA data in both bands with a {\it UV} taper, i.e. setting higher weights to visibilities at short baselines. This was somewhat equivalent to smoothing the maps to a Gaussian beam of 450\,mas FWHM. The map of Band 6 shows that only the shortest baselines were able to detect the emission coming from SW\,Clump. The emission seen in the {\it UV}-tapered map in Fig.\,\ref{fig-allMaps} is exactly as expected for a millimeter counterpart of the visual manifestation of SW\,Clump. The presence of the SW\,Clump emission in Band 7 data is less convincing. The extended emission is missing owing to a different {\it UV} coverage. The ALMA Band\,7 data combined with the SMA survey data, also shown in Fig.\,\ref{fig-allMaps}, have a better {\it UV} coverage but the SMA data did not have enough sensitivity to detect the extended emission of SW\,Clump. 

\subsection{Spectral index}    
\begin{figure}
\includegraphics[trim=20 20 60 50, clip, width=0.5\textwidth]{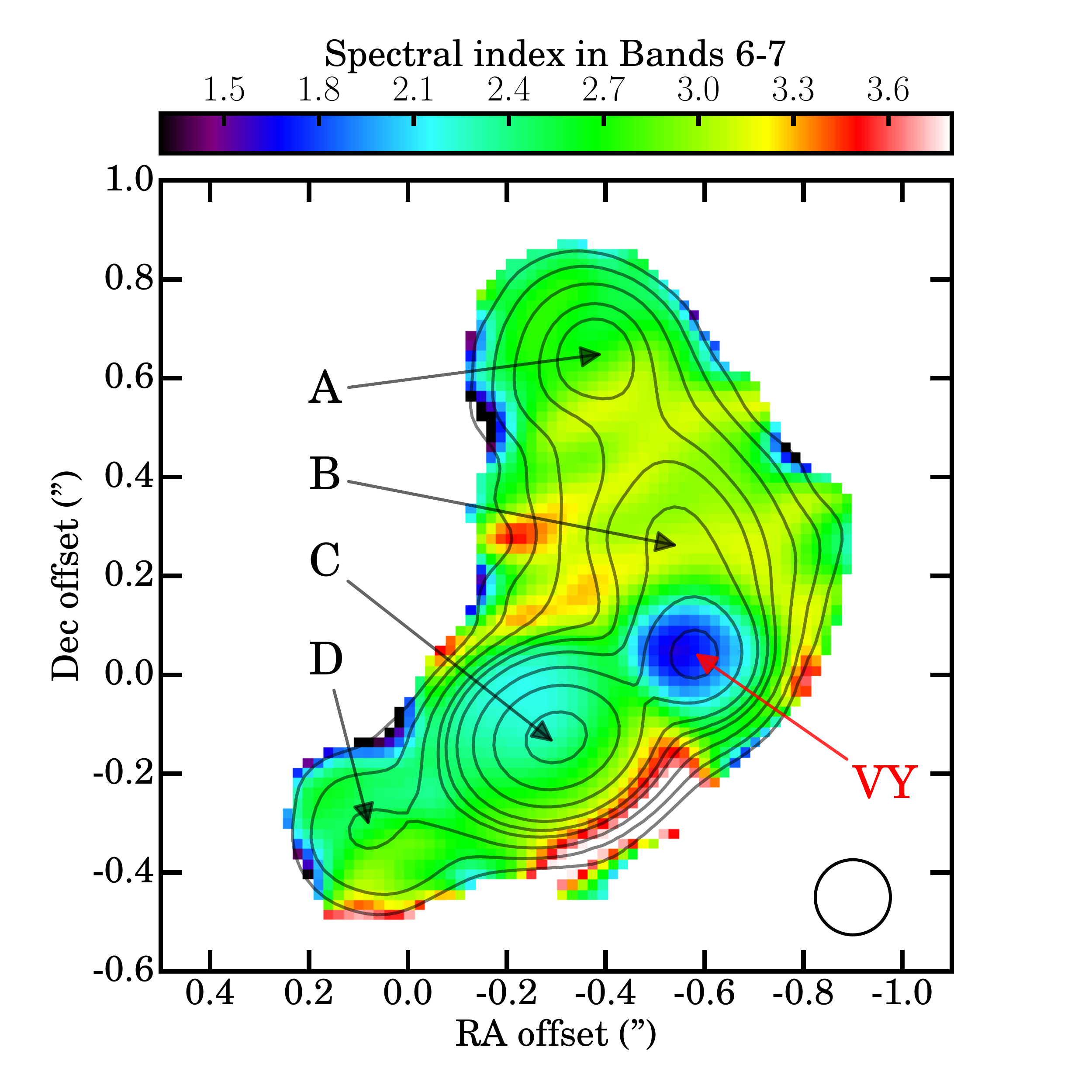}
\caption{Spectral index map of the continuum emission of VY\,CMa between about 230.3 and 351.4\,GHz. Contours represent continuum emission in Band\,6 at 0.5 ($\approx$5$\times$rms of the map noise), 1, 2, 4, 6, 10, 20, 40, 80\% of the peak flux of 95\,mJy/beam. The restoring beam size is shown in the lower right corner. The naming convention for the clumps is indicated by the labels and arrows. At 1.17\,kpc, 100\,mas corresponds to 117\,AU.}\label{fig-spindex}
\end{figure}

We were able to generate a map of the spectral index by comparing the Band 6 and 7 data. Their central frequencies are at 230.3 and 351.4\,GHz. By spectral index $\alpha$, we understand here the exponent of flux density as a function of frequency, $F_{\nu}\propto \nu^{\alpha}$. The spectral-index map was produced from maps restored with a circular beam of 152\,mas, which is a geometric mean FWHM of the beam size of Band\,6 observations. Maps were also shifted to match their positions of the brightest emission peaks.  The map is shown in Fig.\,\ref{fig-spindex}. Typical random 1$\sigma$ errors in the spectral index in the main emission regions is of 0.1 but near the emission peaks it is as low as 0.01. 

The typical spectral index of the overall extended emission of clumps A to D, excluding VY, is 2.8. That figure is consistent with earlier observations at mm and submm wavelengths \citep{OGorman,Vlemmings}. Clump VY is most conspicuous through a very low  $\alpha$ of 1.6$\pm$0.1. We observe a gradient in $\alpha$ across clumps VY and C, which is most likely related to variations in optical depth. 



\subsection{Relation to gas tracers}
Molecular gas observed with ALMA simultaneously with the continuum shows much more complex and extended spatial distribution within the nebula than the continuum. None of the spatial structures, except for SW\,Clump \citep{kamiSurv,DecinNaCl,elvire}, can be uniquely identified as a counterpart of any of the major continuum clumps. In particular, as noted before by \citet{OGorman}, molecular emission seems to ``avoid'' regions of strong continuum emission. We confirm that the emission appears to be weaker toward the main clumps compared to the large-scale distribution. This effect is most likely related to absorption of continuum and molecular self-absorption along the line of sight. Temperatures of this gas must be lower than that of the `photosphere' of a given clump. This is most obvious for clump C which is marginally resolved. Toward clump C, we observe pure net absorption spectrum of several molecules (see Sect.\,\ref{sec-C}). In an attempt to better constrain the location of each continuum clump along the line of sight, we investigated the radial velocities of gas that is potentially associated with the spatial structures seen in continuum emission.   

The ALMA data and earlier mm and submm and visual observations show spectral lines with maximal width of $\sim$120\,\kms\ near the base. This indicates an overall outflow terminal velocity of about 60\,\kms\ (or 10.8\,mas\,yr$^{-1}$ at 1.17\,kpc). We adopt this value as the outflow velocity in the VY\,CMa's envelope although it may not represent true expansion velocities of the individual structures, especially those closest to the star. The uncertainty in the expansion velocities is a consequence of particularly complex kinematics of the inner envelope of VY\,CMa: ({\it i}) observations in atomic lines suggest that discrete structures may be moving with different (terminal) velocities \citep{H2007}; ({\it ii}) maser observations indicate non-isotropic velocity gradients with a typical increase in expansion velocity from 8 to 35\,\kms\ between 75 and 440\,mas radius from the star \citep{anita98} but which not necessarily continue to larger radii \citep{anita}; and ({\it iii}) some gas structures located within tens of arcsec from the star may be falling back on it \citep{zhang}. However, in the framework of a radiatively-driven wind, the dusty clumps we analyze here should be driven directly by radiation pressure, move radially outwards, and reach the terminal velocity closer to the star than the gaseous component. The adopted 60\,\kms\ is certainly a good observation-based upper limit on the terminal and expansion velocities. The stellar center-of-mass LSR velocity is of about 22\,\kms\ \citep{kamiSurv}. 

We extracted and examined molecular spectra of several species (e.g., CO 3--2 and HCO$^+$ 4--3) in apertures placed at the location of the main continuum clumps. In the second-to-last column of Table\,\ref{tab-obs}, we list the ranges of LSR radial velocities of gas seen in emission or absorption toward the lines of sight of the clumps. The ranges are very broad and encompass both red- and blue-shifted gas for all of the features. The line profiles have very complex substructures for all regions and there is no straightforward way to associate any of the velocity components with the dust clump. None of the clumps appears to be significantly blue-shifted, though. 

The extended SW\,Clump has been relatively well characterized in molecular and atomic gas \citep{kamiSurv,H2005,DecinNaCl}. Its CO spectrum displays at least four distinct velocity components centered at about --16, 18, 38, and 53\,\kms. The second component is strongest, broadest (FWHM=17.2\,\kms), and its spatial characteristics matches relatively well to what is seen in scattered-light images from the HST. This indicates that SW\,Clump  has nearly the same radial velocity as the star and thus is moving nearly in the sky plane. This location has indeed been suggested in earlier studies of SW\,Clump \citep{H2007} and is supported by its high polarization at visual and infrared wavelengths (should it be  entirely located in the blue-shifted part of the outflow, no strong scattering and polarization would be expected).   

Following the case of SW\,Clump, we examined the central velocity of the strongest CO emission component seen in the spectra and list them in the last column of Table\,\ref{tab-obs}. If they represent the real gas components of these features, most of the dusty clumps must be located in or near the plane of the sky. Clumps A and B would then be slightly blue-shifted and placed in the near side of the envelope. This is however inconsistent with the visual total-power and polarization data of \citet{peter} as these clumps are not visible in the scattered and polarized light contrary to what would be expected under the assumption of single scattering. The association of the dust clumps with gas remains therefore inconclusive. 

The difficulty in associating dust and gas features may indicate that the examined gas tracers do not show any clear abundance or excitation enhancement within the dusty submm clumps. It is possible that the two phases are decoupled and the dust-to-gas mass ratio within the dusty clumps is significantly higher than commonly assumed for circumstellar envelopes.  Perhaps observations at even higher angular resolutions would help in unique identification of the associated gaseous components and radial velocities of the dusty clumps.

\subsection{Resolved observations of clump\,C} \label{sec-C}
The bright clump C located southeast from VY\,CMa was described in \citet{OGorman} \citep[see also][]{Vlemmings} and was found to be optically thick throughout the mm--submm region (183--690 GHz). This is also consistent with our modeling presented Sect.\,\ref{sec-model}. Measurements of the extent of the feature based on the assumption of an intrinsic Gaussian shape yield a FWHM size of 181$\times$97 mas which is comparable to the angular resolutions of our observations. For optically thick emission filling the beam, one can calculate a brightness temperature which should be close to the dust temperature at the clump's `photosphere'. Assuming a beam-filling factor of unity (implying no substructure within clump C), the peak flux in Band\,7 translates to a brightness temperature of 163\,K. The dust in the clump is cooler than most of the molecular material at similar projected distances from the star (typically $T_{\rm ex}$>340\,K; Kami\'nski et al., in prep.). This may suggest thermal decoupling between gas and dust but it is also possible that we trace no molecular signatures of clump C in the ALMA data.


We observe many species in pure absorption toward clump C. This is especially apparent in Band\,7 data which have the highest angular resolution among all ALMA data published so far. The absorption spectrum is produced in gas of an excitation temperature that is lower than the brightness temperature of clump C. In order to investigate the properties of the gas seen in absorption, we extracted spectra at the position where most molecules show minimum flux. At this position, the brightness temperature is 137\,K. 

Pure absorption is seen in lines of SO$_2$ and NS. Mixed absorption and emission profiles are seen in lines of SO and in CO 3--2. All features are assigned to transitions at $\varv$=0. We simulated the absorption components assuming LTE conditions and using CASSIS\footnote{\url{http://cassis.irap.omp.eu}}. Single excitation temperature was assumed for all species and best agreement with observations was found at $T_{\rm ex}$=52\,K (3$\sigma$ range of 37--67\,K). This temperature is indeed much below the continuum brightness temperature. The absorption components were represented by Gaussian profiles with a FWHM of $\lesssim$10.7\,\kms\ but the observed features, especially those of SO and CO, have more elaborate substructure indicative of multiple -- mostly unresolved -- components. Typical LSR velocity of the absorption features of $\sim$--5.5\,\kms\ indicates that the gas is moderately blue-shifted, i.e. by 27\,\kms\ with respect to the systemic velocity. (Column densities of the different species were also derived but are irrelevant for the discussion here). There is no evidence that this gas is part of clump C.    



In the same spectrum, we see pure emission features of CO, SO, SiS, and SiO arising in gas of higher excitation ($T_{\rm ex}\!\gtrsim$300\,K) and at LSR velocities of +16 and 43\,\kms. They represent regions along the line of sight different than the one seen in absorption features. The spatial characteristics of these emission components rather exclude their physical association with clump C. However, the presence of the component which is red-shifted with respect to the systemic velocity by about 21\,\kms\ may be important for constraining the location of clump C along the line of sight. Since clump C is optically thick, nothing behind it could be seen and thus the material seen in emission must be located between us and the clump. If gas is ejected radially from the star, redshifted gas can be seen only in the far side of the envelope. This would place Clump C in the far side of the envelope as only this location would allow us to see both red-shifted and blue-shifted gas along its line of sight. However, since clump C is only marginally resolved (i.e., the source size is comparable to the beam), the spectrum analyzed here could be contaminated by extended emission surrounding clump C and only observations at even higher angular resolutions could led to more conclusive location of C based on the radial velocity.

\section{3D radiative transfer model of the dusty envelope}\label{sec-model}
To gain a better insight into the properties of the dusty clumps seen by ALMA, we performed 3D simulations of the environment in the RADMC-3D
 code \citep{radmc3d} version 0.41 and with python analysis routines radmc3dPy\footnote{\url{https://www.ast.cam.ac.uk/~juhasz/radmc3dPyDoc/index.html}}. The stellar radiation was represented by a synthetic spectrum with $T_{\rm eff}$=3500\,K, solar metallicity, and $\log g$=--0.5 (lowest available\footnote{at \url{http://phoenix.astro.physik.uni-goettingen.de}}) and generated with the PHOENIX model atmospheres \citep{phoenix} up to 5.5\,$\um$m. At longer wavelengths, we used a Planck function  of the same temperature. The full spectrum was scaled to give a bolometric luminosity of 2.7$\cdot$10$^5$\,L$_{\odot}$. These stellar parameters are consistent with estimates of \cite{wittkowski}. We adopted a distance of 1.2\,kpc \citep{zhang}.

The mineralogy of the dust in the region of interest is very uncertain \citep{Harwit} but the O-rich composition of the VY\,CMa's photosphere and circumstellar gas implies that the dust is predominantly inorganic. Additionally, the presence of a strong "silicate feature" near 10\,$\mu$m in the spectrum of VY\,CMa suggests that silicates are the main opacity source in continuum. We used opacity and scattering properties of the so-called astro-silicates \citep{DL84} generated in RADMC-3D from the optical constants distributed with DUSTY\footnote{\url{https://github.com/ivezic/dusty}}. We combined dust properties for grains in the size range $a$=0.01--1.0\,$\um$m, with grain number populations given as $a^{-3.5}$, and average dust density of 3.5 g\,cm$^{-3}$, as commonly assumed for interstellar and circumstellar media. Similar dust properties were adopted in the models of \citet{gordon} and \citet{Shenoy2016} except that they included pure metallic iron as an equally relevant opacity source as the silicates. Since recent studies strongly indicate that pure iron dust is very unlikely to be present in circumstellar media \citep{Kimura}, and is certainly absent if grains as large as 1\,$\um$m are considered, we did not include pure iron dust in the model. Our range of grain sizes contains particularly large dust because polarized-light studies imply that significant amount of large grains are responsible for scattering in the visual, at least at distances larger than $\sim$0\farcs3 from the star \citep{peter}. One shortcoming of our model is that it does not account for spatial changes in the grain properties. Such changes are expected, for instance, in grain sizes if the grains grow while increasing distance from the star. We however remove from the model all grains warmer than 1700\,K as no such grains are expected to survive. (Silicate dust, specifically, has an even lower sublimation temperature of 1000\,K.) Scattering phase function was assumed to follow the Henyey-Greenstein approximation. These assumptions yield dust opacity of 0.40 and 0.17 cm$^2$\,g$^{-1}$ at ALMA Bands 7 and 6, respectively. 

In the main simulation, we modeled a Cartesian cube of a size of 2000\,AU with 150 grid points in each direction. The cell size of 13.3\,AU is comparable to the size of the star; the latter was represented by a point source. Clumps A to D were simulated as 3D structures of a Gaussian density profile. The choice of a Gaussian distribution was dictated only by the simplicity of the implementation. Whereas observations constrain the clump sizes and overall relative location in two spatial dimensions, the extent of a given feature along the line of sight was a free parameter. We included scattering in the radiative transfer calculations. It turned out early on in the simulations that the dusty medium is very optically thick for visual and ultraviolet (UV) photons making the calculations particularly time-consuming. To speed them up, we used Modified Random Walk method implemented in RADMC-3D \citep{RandomWalk}. The number of photons used in the Monte Carlo simulations varied from $10^5$ to $10^9$. For each spatial configuration of dust distribution, first a thermal structure was calculated. Based on that, regions where dust temperature exceeded the sublimation temperature were removed from the simulation and the thermal structure was calculated once again. With that, we reconstructed images which were further processed in CASA for a direct comparison to ALMA continuum maps. Given a very high number of free parameters (over 40 in the basic version of the simulation), best configurations were searched through iterative trial-and-error modifications of the models. We aimed to construct only a general 3D model of the dusty environment requiring a minimum dust mass possible. We attempted to reproduce the peak and integrated ALMA fluxes of each feature within the order of magnitude. This aim, however unambitious it may appear, was challenging without increasing further the number of free parameters. For example, better fits to total flux and its distribution would require considering other than Gaussian density distributions. 
  
Constraining the relative locations of the clumps in 3D is a complex task, especially because most of the clumps are very optically thick to UV and visual radiation. Shadowing, including self-shadowing, was perhaps the foremost problem complicating the modeling. For instance, projected in the sky, cloud D appears to be a continuation of cloud C toward east. Similarly cloud A may be considered a continuation of feature B toward north. However, features C and B produce significant and very extended shades. At the projected distance of cloud D, clump C produces a shade of a radius of 90\,AU (and its penumbra is even larger). For complexes A--B and C--D, to each form one radially extended feature, clumps A and D would have to be located in the shadows of B and C, respectively. In such configurations, B and D require unreasonably large amounts of dust to explain their ALMA fluxes. We therefore located them in unshadowed regions, far from all other clumps. Another interesting effect complicating the simulations is that scattering of visual light in this complex 3D configuration allowed illumination of some clumps from many angles, not only directly from the star, affecting the thermal structure of the dusty medium. As the result, the apparent intensity, position, and shape of the structures are strongly coupled to their relative locations. Most configurations are very unstable to even small changes and it is very difficult, if not practically impossible, to explore a wide range of free parameters of the model. 
 
In particular, in all models of clump C that we considered, the feature is massive and very optically thick. We required an unreasonably high mass of dust -- of the order of 1\,M$_{\sun}$ -- to roughly approach the observed fluxes and extent of clump C. The large size of the clump in radial direction makes it very optically thick for the stellar radiation and self-shielding of the far part of the clump (as seen from the star) is very effective. In our basic implementation of the feature, most dust is cool and dark adding to the clump's total mass but not to the observed mm and submm fluxes. These problems can be partially explained by too low opacity adopted in our simulations or by disregarding  substructures within the clump. We suspect the latter is more important. Cool stellar winds show substructures at different scales, including porosity \citep[see][and references therein]{porosity}. We implemented basic porosity in clump C in order to explore its effect on submm fluxes. For example, instead of a single Gaussian, we simulated clump C as five Gaussian substructures with a three times smaller FWHM, and spread evenly within the same volume as the single component. Stellar radiation could penetrate deeper into the clump in such a configuration increasing the total submm fluxes of clump C by a factor of 2.2 at the same mass of dust involved. Some level of porosity therefore can help explaining the observations with a considerably lower dust mass. We are convinced that clump C has an unresolved substructure but its characteristics (e.g. density contrast or geometry) cannot be currently constrained. By ignoring the porosity in simulations of other clumps, we may be overestimating their masses, too. It is however impractical to perform a full simulation of this kind for all clumps as it would increase the number of free parameters even further. For these reasons and other problems mentioned above, we treat the radiative transfer modeling performed here only as a qualitative toy model aimed to explore the possible geometries of the system. We hope it to pave the way for more elaborate models of VY\,CMa's circumstellar environment. 

\begin{figure}
\includegraphics[angle=0,width=0.49\columnwidth,trim={0 0 0 0},clip]{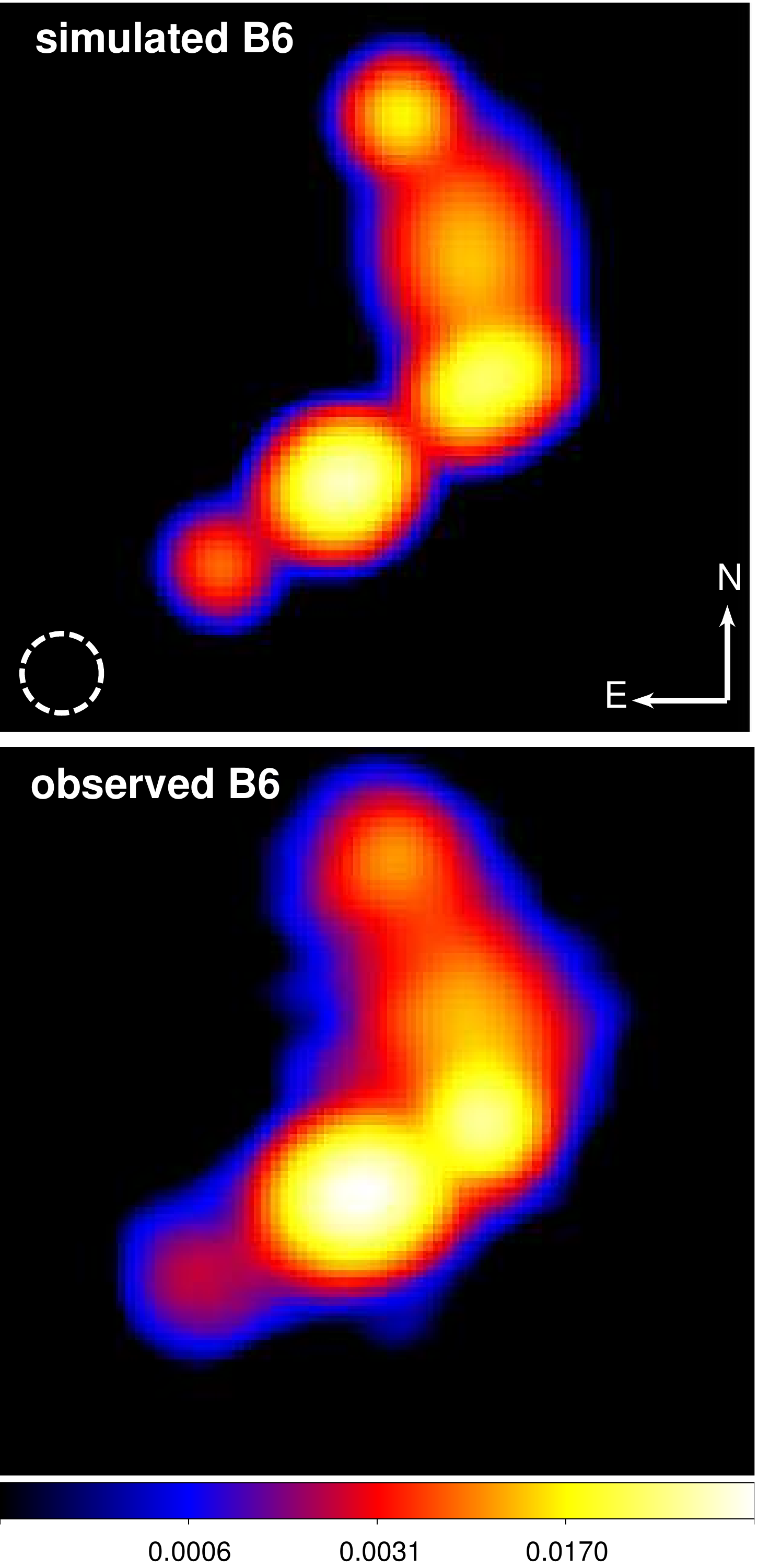}
\includegraphics[angle=0,width=0.49\columnwidth,trim={0 0 0 0},clip]{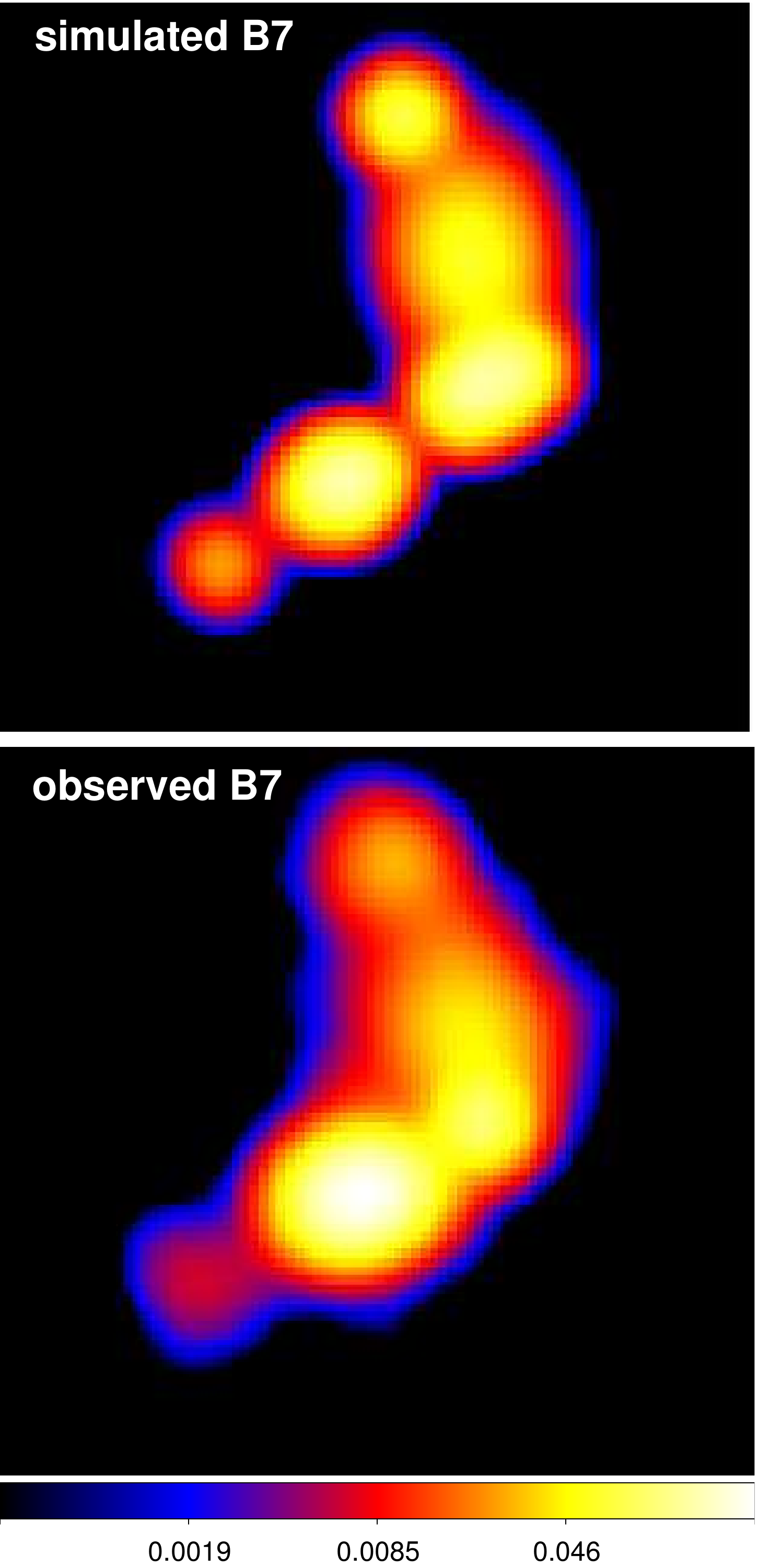}
\caption{Simulated and observed maps of continuum emission of VY\,CMa. All images are in scale and were generated with a beam of 152\,mas FWHM (shown with the dashed line). Observed and simulated images in each band are displayed with the same logarithmic color scale (in Jy/beam).}\label{fig-sim}
\end{figure}

\begin{figure*}
\includegraphics[angle=0,trim={0 0 0 0},clip,width=1.0\textwidth]{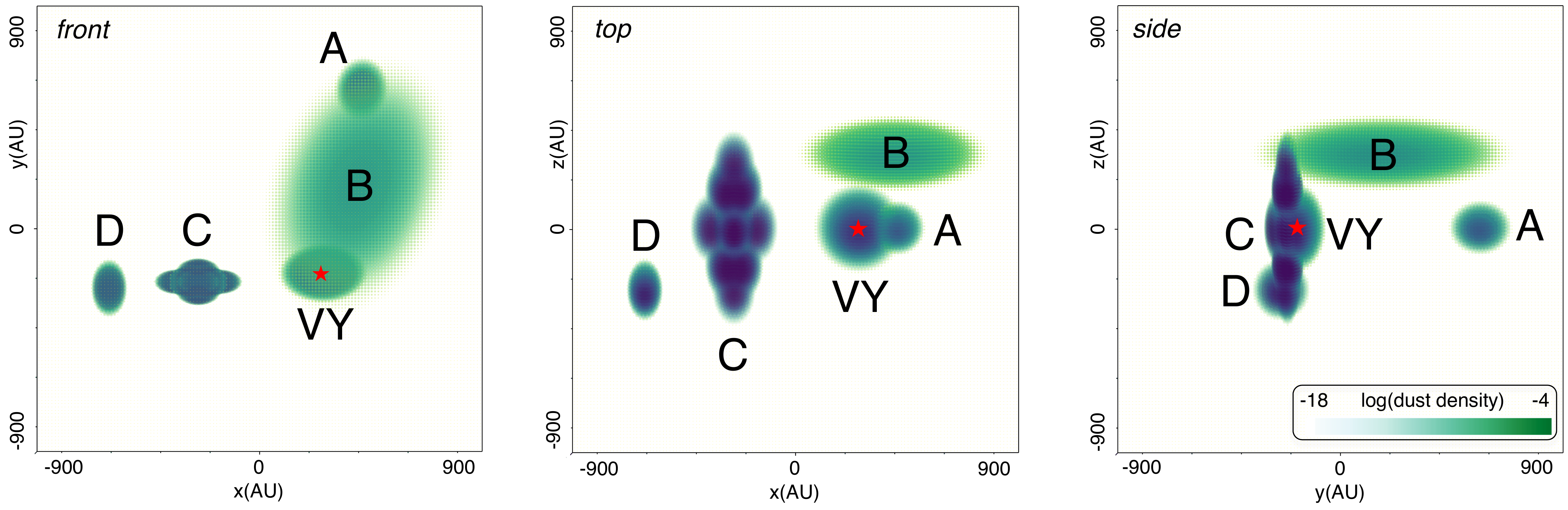}
\caption{Rendering of the 3D model of dust distribution near VY\,CMa. The cube of 2000\,AU is shown from three different perspectives (see labels). The star is located at ($x,y,z$)=(250,--186,0)\,AU and marked with the star symbol. Colors represent density an at arbitrary surface.
}\label{fig-3D}
\end{figure*}

From about 1000 simulations performed, we choose one that relatively closely reproduces the ALMA observations with a total dust mass of 0.47\,\msun, 98\% of which resides in problematic clump C. Our preferred 3D model structure is shown in Fig.\,\ref{fig-3D}. Simulated observations are illustrated in Fig.\,\ref{fig-sim}. Table\,\ref{tab-sim} characterizes the discrete clumps in a model of our choice but this characteristics by no means represents unique parameters of these structures. Ages were calculated from the modeled positions assuming a velocity of 60\,\kms. Below, we comment on individual clumps considered in the model.

\begin{table*}
\caption{Clump parameters used in radiative-transfer simulations.}\label{tab-sim}
\centering
\small
\begin{tabular}{c | ccc | c | ccc | c | cc}
\hline
Clump& \multicolumn{3}{c|}{Distance from the star (AU)} & Age & \multicolumn{3}{c|}{FWHM size (AU)}   & Dust mass &  \multicolumn{2}{c}{Max opt. depth} \\
ID   & $x$  & $y$  & $z$                           & (yr)& $x$  & $y$  & $z$& (\msun)   &  $\tau_{\rm B6}$ & $\tau_{\rm B7}$ \\
\hline
VY	 &   0 &   0  &  0  &<20 &  90 &  90 &  90 & 0.014 & 0.03  & 0.07 \\ 
A	 & 155 &  745 &  0  & 60 &  60 &  70 &  60 & 0.002 & 0.10 & 0.24 \\ 
B	 & 140 &  350 &300  & 38 & 230 & 350 & 100 & 0.004 & 0.05 & 0.11 \\ 
C	 &--500&--35  &  0  & 40 & 120\tablefootmark{a} &  60\tablefootmark{a} & 200\tablefootmark{a} & 0.46 & 18.1 & 42.0 \\ 
D	 &--865&--60  &--250& 71 &  34 &  56 &  60 & 0.002 & 0.10 & 0.22 \\ 
SW  &   150&--1300&  0  &104 & 400 & 160 & 240 & 0.001\tablefootmark{b} & 0.01 & 0.03 \\
\hline
\end{tabular}
\tablefoot{
Distances and sizes are given in a Cartesian system, $xyz$, where negative $z$ values represent the near side of the envelope. Masses are overestimated (see text). Maximum optical depths were determined in images simulated with a beam of 135\,mas.
\tablefoottext{a}{Sizes given for a single-Gaussian implementation of the feature rather than the multi-component one described in the text.}
\tablefoottext{b}{Based on observations with a missing-flux problem.}
}\end{table*}

\subsection{Feature VY}\label{sec-vy}
Feature VY may not be a dusty clump in the same sense as the other features since it has a significant flux contribution from the star. In our model, VY is modeled as dusty medium of a Gaussian density distribution centered on VY\,CMa and uniformly surrounding the star, as in an idealized spherical outflow. Alternatively, however, it could be a clump located in front or behind the star and seen along the same line of sight. In our implementation, we clear of dust the central space within the radius of 90\,AU (equivalent to a diameter of 0\farcs15) where solids would be too warm to exist heated by the cool but very luminous star. The peak surface brightness simulated with a beam of 152\,mas are 34 and 140 mJy/beam in Bands 6 and 7, respectively. Of that, 13.0 and 30.1\,mJy/beam comes from the black body radiation of the model star. \citet{OGorman} noted that stellar fluxes are underestimated in a black-body model at mm and submm wavelengths owing of the unaccounted presence of an extended radiophotosphere. We extrapolated fluxes expected from their radiophotosphere model getting 12\,mJy in Band\,6 and 44\,mJy in Band\,7. Given the small difference with our assumed black-body fluxes and uncertainties in the radiophotosphere model of O'Gorman et al., we did not correct our model for this effect. We required 0.014\,\msun\ of dust to explain the observed fluxes which is rather large compared to most other clumps (except C). The stellar contribution partially explains the very low spectral index of 1.7 toward VY because the stellar spectrum is expected to have $\alpha\!\approx\!2$ \citep{lipscy}.  At the modeled parameters of VY, the medium must be very thick at visual light and obscure the stellar photosphere from a direct view for a terrestial observer, as indeed visual observations seem to indicate. The stellar photosphere and its light variations are seen mainly through light scattered in the surrounding medium with a high degree of inhomogeneity \citep{H2005}.

\subsection{Clump A}
Clump A appears as the most distant feature in the north. Because it is not seen in the scattered light or polarization maps, its location in the near envelope is unlikely. To put it as close as possible to the star -- and by that minimize its mass in the simulation -- we located A in the plane of the sky. In our 3D configuration, the scattered light of clump A irradiates the surface of clump C. 

\subsection{Clump B}
Clump B is very likely composed of multiple features and is associated with a weak extended component which is not reproduced in the simulation. This larger region extends 0\farcs8 in the east-west direction. Clump B may be physically connected to VY through a bridge of emission. It may be also physically bridged with feature A, but in our simulation, we tried to locate both clumps far apart so that A is not shadowed by B. In the simulation, we located B 300\,AU behind the sky plane. 

\subsection{Clump C} 
Clump C was identified and characterized in some detail in the earlier studies based on ALMA observations \citep{anita,OGorman,Vlemmings}. Our definition of C is somewhat different, though, because feature D is treated here as a separate knot whereas before it was considered as an extension of C. Clump C is the most massive, largest, and brightest of the analyzed features. We made the clump particularly elongated in the line of sight (see Fig.\,\ref{fig-3D}) to increase its brightness without increasing its size in the two other directions which are constrained by observations. None of our models could explain the observed fluxes with masses lower than 1\,\msun\ and we very likely overestimate the mass also in models that give lower-than-observed fluxes. In our favorite model presented here, the fluxes of C are still about twice lower than observed. We interpret this failure in modeling as a signature of a complex substructure within the clump (Sect.\,\ref{sec-model}). Another scenario that, in principle, could explain the high flux of clump C is that it contains an internal heating source, such as a low-mass star. We however dismiss such a possibility as there is no other evidence supporting the presence of an embedded companion. All earlier claims of a binarity of VY\,CMa have never been confirmed. 

Future observations at higher angular resolutions will verify our suggestion of substructure in clump C. Such observations should preferentially be done at submillimeter wavelengths. In images simulated for infrared and shorter wavelengths and at angular resolutions a few tens of mas, feature C appears as a narrow crescent illuminated by the star and with bulk of its mass being dark in its own shadow.

We locate the clump in the plane of the sky, although spectral observations suggest (Sect.\,\ref{sec-C}) it may be mostly located in the far (redshifted) side of the envelope.

\subsection{Clump D}
In our model, clump D is placed far off the sky plane (250\,AU) to avoid the shadow of clump C (which is assumed to be located closer to the star). Although very weak in the ALMA maps, in certain configurations clump D may contain a significant amount of dust, even comparable to that in the younger clump C. This relatively high mass of D in some models is imposed by the large physical distance of the clump from the star (necessary to avoid the shadow of C) or by a low dust temperature if D is an a thermal shadow of C. 

It is possible that some polarized features seen in the visual maps from SPHERE \citep{peter} are optical manifestations of clump D. We therefore placed it in the near side of the envelope where forward scattering would make it a bright visual source. This location is consistent with the presence of CO emission at negative LSR velocities toward this structure. Note that in our implementation, clump D is considerably older than clumps A to C but still appears relatively compact in our maps and in polarized light. It is unclear whether the dust clumps expand as they move away from the star. 

\subsection{SW Clump}
Southwest (SW) Clump is a discrete structure within the envelope that has been studied most extensively at visual and infrared wavelengths, up to 12\,$\um$m \citep{ShenoySW,gordon}. The structure is located $\sim$1\arcsec\ south-west from the star, is extended ($\sim$11\arcsec), and early observations confused it with a hypothetical companion of VY\,CMa (never confirmed). Currently, it is thought to be a single (or double) clump ejected from the star 300--500 years ago and containing at least 5.4$\cdot$10$^{-5}$\,M$_{\sun}$ of dust (for a review of properties of SW\,Clump, see \citet{Robinson1971} and \citet{gordon}. The clump was found to be optically thick at wavelengths $<$9\,$\mu$m and all its measured fluxes are dominated by scattered light with very uncertain contribution from thermal dust emission. Recently, SED modeling of \citet{gordon} suggested it should be observable with ALMA in thermal emission but earlier observations of the submm continuum did not detect it, even despite a very high sensitivity. 

Our observations are not well suited to characterize the distant and extended SW\,Clump because the lack of sufficiently short baselines affects the surface brightness of the feature in our {\it UV}-tapered maps (Sect.\,\ref{sec-obs}). Nevertheless, we extended the RADMC-3D model to larger distances and implemented SW\,Clump into the density distribution as a single Gaussian clump whose center is located in the plane of the sky. The structure is relatively extended allowing forward-scattering off dust grains in the near side of the clump.  We processed the simulated maps through the baseline filter consistent with {\it UV}-tapered observations in Band\,6 (using the ALMA's {\it simobserve} and {\it simanalyze} tasks). The model yields dust temperatures of $\lesssim$100\,K inside SW\,Clump but the clump's rim, directly illuminated by the star and by the other clumps, is warmer ($\sim$200\,K). These values are consistent with earlier estimates of temperatures, even though they assumed a different dust composition \citep{ShenoySW, gordon}. The observed fluxes in Band\,6 can be roughly reproduced with 1$\cdot\!10^{-3}$\,M$_{\odot}$ of dust which is much higher than 5.4$\cdot\!10^{-5}$\,M$_{\odot}$ estimated by \citet{gordon} based on their SED reconstructed from observations at shorter wavelengths ($\lambda\!<$12\,$\mu$m). Additionally, ours is a mass lower limit since some flux is still missing in the Band\,6 observations. The large discrepancy in masses is expected because mm-wave observations are sensitive exclusively to thermal continuum emission and likely probe the entire dusty clump whereas flux of scattered visual and infrared light is highly depended on the viewing angle and geometry of the source. In our simulated configuration, the visual emission is coming only from the clump's rim which is directly illuminated by the star (and its surroundings) and most of the clump mass is dark at these wavelengths. The physical center of the clump is located further away from the star than it appears at visual HST images. This effect also partially explains the apparent mismatch between the position of the visual emission of SW\,Clump and that seen in our Bands 6 and 7 maps at Briggs weighting (Fig.\,\ref{hst} and left panels of Fig.\,\ref{fig-allMaps}). We did not attempt to reproduce the entire SED of the feature in the current study because the visual images show a complex substructure of SW\,Clump \citep{Smith2001,gordon} which we could not reproduce with our simple model. 


The simulated mm-wave emission is very optically thin with $\tau_{\rm B6}\lesssim$0.01. At visual, where the rim is the only visible signature of the simulated clump, the material is very optically thick with $\tau_{\rm 0.55\mu m}\!\sim$1700, in qualitative agreement with earlier results \citep{ShenoySW, gordon}. The clump becomes optically thin only at wavelengths longer than 175\,$\mu$m but most far-infrared observatories would not be able to spatially resolve the clump. In our simulation, the center of the clump is at a de-projected distance of 1310\,AU which indicates it was created close to 1911. (Note that \citet{H2007} estimate the age to be 500\,yr and \citep{gordon} at $\lesssim$300\,yr which are consistent with our result if it is treated as the lower limit). Our calculations need to be verified with better observations of SW Clump at more compact ALMA antenna configurations and with more advanced modeling that would take into account the substructure of the clump. 

Our characterization of the visually-prominent SW\,Clump is not very different than these of other major submm clumps located closer to the star. SW\,Clump is certainly more extended than all other clumps discussed here, possibly in relation to its advanced age (>100\,yr). 

\section{Clumps relation to stellar light variations}\label{sec-ages}
Our rough constraints on the age of the dust clumps (Tables \ref{tab-obs} and \ref{tab-sim}) indicate that most were created in the times when astronomers had observed VY\,CMa in visual light. We analyzed historical light curves of VY\,CMa in search for signatures that can be associated with the creation of the clumps. One possibility is that the events leading to the massive dust-condensation episodes are associated with physical changes of the star that are manifested in visual light, e.g. by flaring. The creation of dusty clumps could also influence the light variations through enhanced extinction, as for instance commonly observed in R\,CrB stars.

\begin{figure*}
\includegraphics[angle=0,width=\textwidth,trim={0 0 0 0},clip]{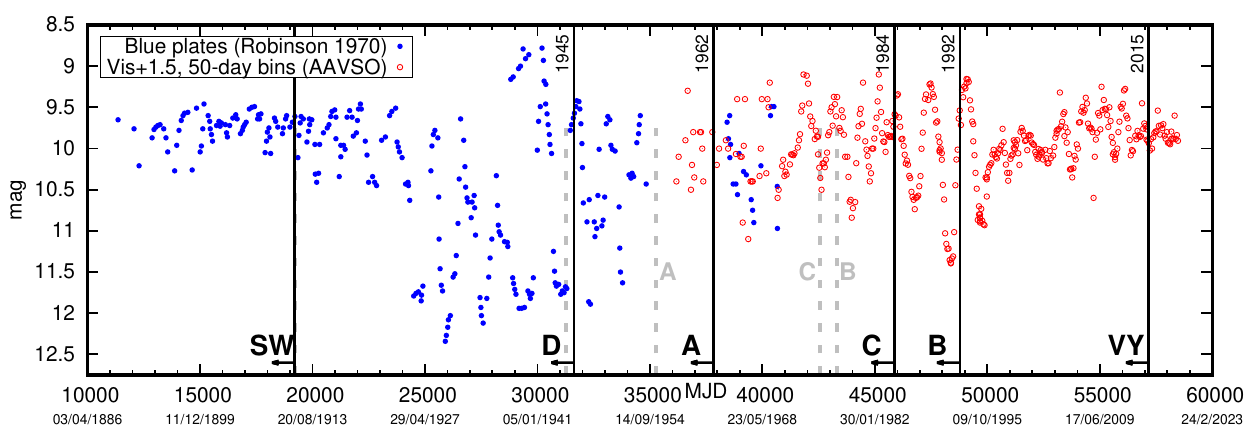}
\caption{Light curve of VY\,CMa. Data come from \citet{Robinson1970} and AAVSO. Measurements are binned to 50 day in both datsets. Verical solid lines mark the approximate lower limits on the creation dates of the major clumps based on their projected distances from the star. The gray dashed lines show kinematic ages that take into account clump locations along the line of sight as proposed in our 3D model.}\label{fig-lc}
\end{figure*}

Visual flux measurements of VY\,CMa have been obtained since at least 1801 and include sparse and uncoordinated measurements over the entire nineteenth century \citep{Robinson1971}. Harvard plates, and other photographic and photoelectric measurements over the period 1893--1971 were processed and compiled by \citet{Robinson1970}. We converted\footnote{\url{https://automeris.io/WebPlotDigitizer}} his {\it blue} light curve to an electronic table from a scan of the original study. A poor quality of the scan limits the accuracy of the reproduced photometry but still suffices our aims here. Regular observations since 1957 have come from AAVSO\footnote{\url{https://www.aavso.org}}. The light curves are reproduced in Fig.\,\ref{fig-lc}. Robinson noticed that VY\,CMa's brightness has been decreasing since the earliest observations. On top of that, there has been long-term variations with a most noticeable drop near 1870. Frequent photometry starting at 1893 shows quiescent periods when the stellar magnitude varies by $\sim$0.5 on time scales of a few hundred days. There were also a few active periods where (unbinned) light variations reached 4\,mag and had a semi-regular period of $\sim$1100 days. Of course, these photometric records illustrate only changes on one hemisphere of the supergiant at the time. Because the stellar rotation is unconstrained, it is unclear how rotation, if present, would influence the observed variations.

As discussed in Sect.\,\ref{sec-vy}, the dust associated with the VY feature is not interpreted here as produced by periods of enhanced mass loss but is rather a signature of steady outflow at a relatively low dust mass-loss rate of 5$\cdot$10$^{-6}$\,\msun\,yr$^{-1}$ \citep{OGorman}. Indeed, the part of the light curve corresponding to the 1--2 decades lifetime of VY is very quiescent. Feature B was born approximately before 1992, possibly even decades earlier. Interestingly, the date 1992 falls in the period where VY\,CMa experienced very strong light variations (3.7 mag). Clump\,C is most likely seen in the sky plane and its real age should be close to our lower limit. This dates its creation to about 1984 when VY\,CMa was moderately active. It is intriguing whether clumps B and C have anything to do with the enhanced stellar variations in 1985--1995. Clump A was born before 1962, possibly decades before that date since it is most likely located in the far side of the envelope. Similarly, Clump D could have been created decades before 1945. The light curve shows that in 1920 VY\,CMa entered a particularly variable  state in which it had remained for a few decades. Clumps A and D could have been created in that phase. Lastly, our estimates show that SW\,Clump must be older than 104\,yr. \citet{gordon} give 300\,yr for its age setting the ejection date close to 1715, before any known photometric records of VY\,CMa. We note that before 1920 and at least since 1893, the star had remained relatively quiescent. The lack of any observable clump' birth dates falling in this time period is consistent with the hypothesis that a dust-forming event is associated with strong light variations. 

Although unique association of the clumps ages with the light curve features is not possible, all observations are consistent with an enhanced stellar variability during or years after the creation of a dust clump. The observed stellar variations are certainly different in nature than those in R\,CrB stars, mainly by the lack of characteristic low-brightness plateaus. Additionally, it seems that dense dusty clumps are rarely formed exactly at the line of sight toward VY\,CMa and fill small solid angles. Consequently, direct substantial obscurations by discrete dust clouds must be rare. Dust condensation might thus occur at large distances, $\gg$3\,R$_{\star}$. Interestingly, Robinson noticed that during the overall century-long dimming, the stellar spectrum becomes redder. This would be consistent with increasing circumstellar extinction. However, physical long-term changes in stellar temperature cannot be excluded for this variable supergiant, as well. As our modeling showed, however, the densest clumps have strong impact on the appearance of the nebula by casting huge light and thermal shadows. This effect can account for the confusing changes in morphology of the nebula observed by different observers over the past century \citep{Robinson1971}.   

\section{Discussion and conclusions} 
Our radiative-transfer model requires 0.5\,\msun\ of dust to explain the observed submillimeter fluxes, yet it still under-predicts by a factor of $\sim$2 the fluxes of clump C. This mass is higher or comparable to the total (dust plus gas) mass estimates in earlier studies of VY\,CMa's envelope \citep[e.g.][]{Shenoy2016,Muller}. Most of the earlier estimates, however, used optically thick tracers (dust, short-wavelength SED, or CO rotational emission) and relied on uncertain conversion factors from dust mass to total mass or from CO mass to total mass. They constitute rough lower limits on the masses of the entire envelope or masses of individual features as in \citet{OGorman} or \citet{gordon}. On the other hand, our masses can be largely overestimated, mainly because the model ignores unresolved substructure within the dusty clumps and dust opacity is essentially unknown (Sect.\,\ref{sec-model}). Moreover, the 3D configuration we proposed is not unique introducing further uncertainties in the mass constrains. 

With the widely used gas-to-dust mass ratio in of $\psi$=200 \citep{vanLoon}, our dust mass constrains would indicate the total mass of the envelope of 100\,\msun\ which is much above the current mass of the supergiant \citep[17\,\msun;][]{wittkowski} and therefore is unacceptably high. The value of $\psi$ is however highly uncertain and essentially unknown for the extreme mass loss episodes of VY\,CMa. Most estimates of $\psi$ have been done for entire circumstellar envelopes -- not discrete features that are so characteristic of VY\,CMa. It is possible that the dusty clumps that we observe with ALMA are dominated by dust with little gas content. Interferometric maps of molecular emission indeed do not show any spatial features that can be uniquely associated with the dusty structures (except for the distant SW\,Clump). Also, there are bright molecular structures that do not have a detectable counterpart in continuum (Kami\'nski et al., in prep.). Low gas abundance in the dusty clumps would make the total mass of the clumps closer to a more reasonable number but it is difficult to accept that $\psi$ would be of the order of, say, 1, which would bring it closer to earlier mass estimates. The modeled masses here should be thus treated with a lot of caution, even though masses of clumps A, B, D, and SW seem more reasonable than that of C.

If we accept for a moment that the simulations presented here represent in some aspects the real physical configuration of the dusty envelope, most of the submm clumps are very optically thick at UV and visual (some even at submm) wavelengths, are dominated by dust (i.e. have very low gas-to-dust mass ratios), and thus are indeed unparalleled signatures of extreme dust formation and mass loss among all known non-explosive stars. With no fine substructure (no porosity), the shadowed parts of the clumps are not exposed to the stellar radiation and therefore do not receive momentum from stellar photons. In such a scenario, these dark parts of the envelope could not have been accelerated in the mechanism that is commonly accepted to act in dust-driven winds of AGB stars \citep{review}. Had these clumps been accelerated before the dust condensation occurred, i.e. when all material was still in gas phase? There is no known mechanism that could accelerate cool (3200\,K) gas from the photosphere to the part of the envelope where we observe the clumps today. Since most of the mass of the VY\,CMa's envelope was lost in such discrete features, these unsolved problems lead to the conclusion that we know remotely nothing about extreme mass-loss mechanisms in red supergiants.


Many studies have suggested that the enhanced mass-loss events indicated by the dusty clumps must be related to phenomena at the stellar surface,  such as extreme convection, pulsation, or magnetic fields. These suggestions are often equivocal and no qualitative description have been proposed. Winds of AGB stars seem in this context much better understood \citep{review,HF2019}. The rather ambiguous relation of the expected birth dates of the main clumps with the visual light curve of the supergiant increases the enigma as it may suggest that nothing particularly extreme is taking place on the stellar surface --- at least when observed in visual light --- when the star is busy creating these dense dusty puffs or their gaseous progenitors. Convective cells of a life span of up to years are seen on surface of Betelgeuse monitored in polarized light but their link to mass-loss episodes in red supergiants has never been well established \citep{specpolBetel}. Interestingly, \citet{Kervella} tentatively proposed that the focused mass ejections present in the circumstellar environment of Betelgeuse originate from long-lived ($>$6\,yr) hot spots seen at the photosphere of the supergiant at multiple wavelengths. Such a potential link cannot be easily verified for VY\,CMa which is several times more distant and whose photosphere has never been resolved.  





The high clump masses we derived certainly require verification in a more advanced modeling approach but suggest that in case of the VY\,CMa's envelope instead of observing an outflow or wind, we actually see chunks of the star being ejected into the interstellar medium. A future quantitative model should be based on observations at an even higher angular resolution that could constrain the form of substructures within the clumps. Constraining the rate and origin of the mass loss of VY\,CMa is essential for our understanding evolution of high-mass stars and many supernovae.

\begin{acknowledgements}
This paper makes use of the following ALMA data: ADS/JAO.ALMA\#2013.1.00156.S. ALMA is a partnership of ESO (representing its member states), NSF (USA) and NINS (Japan), together with NRC (Canada) and NSC and ASIAA (Taiwan) and KASI (Republic of Korea), in cooperation with the Republic of Chile. The Joint ALMA Observatory is operated by ESO, AUI/NRAO and NAOJ. The National Radio Astronomy Observatory is a facility of the National Science Foundation operated under cooperative agreement by Associated Universities, Inc. We acknowledge with thanks the variable star observations from the AAVSO International Database contributed by observers worldwide and used in this research. The 2017 SMA observations were conducted within the the AY191 practical course at Harvard Collage and we acknowledge the participation of N. Patel, Q. Zhang, B. Brzycki, N. James, S. Mehra, and C. Husic in the program. 

I am grateful to K. Menten for discussions and the encouragement to perform the study and to P. Scicluna for making available to me his SPHERE observations. I also thank K. Wong, P. Scicluna, and the referee, A. Richards, for insightful comments on an earlier version of the paper.
\end{acknowledgements}

\end{document}